\documentclass[twocolumn,showpacs,preprintnumbers,amsmath,amssymb]{revtex4}
\usepackage{graphicx}% Include figure files
\usepackage{dcolumn}% Align table columns on decimal point
\usepackage{bm}% bold math

\def\ket#1{\mid #1\rangle}

\begin{document}

\title{Shot Noise for Entangled Electrons with Berry Phase}

\author{Hui Zhao$^{1,2}$, Xuean Zhao$^{1}$, and You-Quan Li$^{1}$}

\address{$^{1}$Zhejiang Institute of Modern Physics, Zhejiang University, Hangzhou 310027, China\\
$^{2}$Department of Physics, Anshan Normal University, Anshan
114005, China}

\date{December 29, 2004}

\begin{abstract}
We study shot noise for entangled electrons in a 4-lead
beam-splitter with one incoming lead driven by adiabatically
rotating magnetic fields. We propose a setup of an adiabatically
rotating magnetic field therefor, which is appropriate for an
electron beam to transport through. Using the scattering matrix
approach, we find that shot noise for the singlet and that for the
entangled triplet oscillates between bunching and antibunching due
to the influence of the Berry phase. It provides us a new
approach for testing the Berry phase in electron transport on the
basis of entanglement.
\end{abstract}


%\keywords{Suggested keywords} %

\maketitle

Since the quantum interference and quantum entanglement in
electron transport become very important, the study of biparticle
system has absorbed much attention recently. The extension of the
scattering matrix approach \cite{scattertheory} to deal with a
biparticle system in a beam-splitter \cite{beamsplitter,
beamsplitter2} is instructive \cite{setup}. Within the extension,
entangled states are identified via shot noise measurement. Some
application of the extended approach involves dealing with
entangled states affected by an interaction, such as the Rashba
interaction \cite{rashba}. Because the interaction affects
entangled states, it will finally affect the transport current and
shot noise which explicitly involve the effects of interaction.
This provide us an approach to probe the properties of the
interaction. As to the Berry phase \cite{berryphase}, it is
studied very well theoretically and experimentally \cite{book} in
a single-particle system. However, the interest in the Berry phase
of a biparticle state has just been aroused since the work of
Sj\"{o}qvist \cite{sjoqvist}. Two aspects of the Berry phase for a
biparticle system are very important. One is how the Berry phase
changes due to the interparticle interaction \cite{sjoqvist,
hessmo, tong}, the other is how it affects the biparticle state
\cite{spinecho1}. All those advancements lead us to a way to
observe the Berry phase in electron transport with an entangled
state driven by adiabatically rotating magnetic fields.

In this Letter, we study the transport properties of entangled
electrons in a beam-splitter configuration \cite{beamsplitter,
beamsplitter2} with an incoming lead driven by adiabatically
rotating magnetic fields. We propose a setup which is appropriate
for an electron beam to transport through therefor. The Berry
phase affects the spin states of a single electron significantly.
The application of this set up can make entangled state into a
superposition of the singlet and triplets.
We calculate shot noise for
one outgoing lead by means of scattering
matrix approach \cite{scattertheory}
We find the Berry phase affects the entangled
state and shot noise for the mutated state oscillates as a function
of the Berry phase, which helps us to observe the Berry phase in
electron transport on the basis of entanglement.

%Thus the electrons travelling along the axis of the cylinder is expected to
%feel a right-hand rotating magnetic field.
%
\begin{figure}[bph]
%\vbox to1.5in{\rule{0pt}{2in}} \special{psfile=fig_1.prn
\vbox to2.5in{\rule{0pt}{2.4in}} \includegraphics{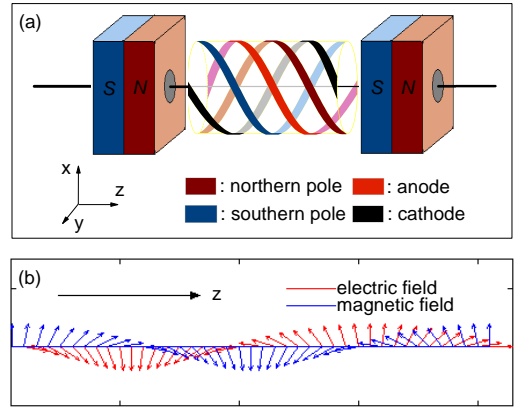}
\caption{Schematic view of a design of an adiabatically rotating
magnetic field. (a): the design of an adiabatically rotating
magnetic field, which is appropriate for an electron beam to
transport through it. It consists of two magnets to produce a
constant magnetic field component along the $z$ axis, two magnetic
stripes to produce a helically radial magnetic field component and
two metal stripes to produce a radial electric field. (b): the
configuration of the radial magnetic field component and the
radial electric field.} \label{fig1}
\end{figure}

The conventional adiabatically rotating magnetic field is
considered around a fixed point, but it is not suitable for an
electron to transport. We propose a setup which is suitable for an
electron beam to transport through. The setup consists of a
cylinder, two magnets, two magnetic stripes and two metal stripes.
The cylinder is displaced between the two magnets, around whose
surface the two magnetic stripes and the two metal stripes are
wound helically, as shown in Fig.~\ref{fig1} schematically. Our
purpose is to make electrons travelling along the axis of the
cylinder feel a right-handed rotating magnetic field. Thus we need
to realize the two components of the rotating magnetic field as
follows. The constant axial component $\vec B_z$ is produced by
the two magnets. Each magnet is made with a hole coaxial to the
cylinder, so that electrons can enter or leave the cylinder
through the hole. In our setup, the radial component $\vec B_\bot$
distributes uniformly but rotates helically along the axis of the
cylinder, which is produced in the following way. Starting from
two opposite positions at one end of the cylinder, we wind two
magnetic stripes around its surface respectively with a right-hand
uniform helix such that they undergo a twist of $2\pi$ when
arriving at the other end. We keep the inward poles of the two
magnetic stripes opposite. These two magnetized stripes clearly
produce the wanted component $\vec B_\bot$. However, when an
electron moves along the axis of the cylinder, it will suffer a
Lorentz force arising from this component. The Lorentz force will
make the moving electron deviate away from the axis of the
cylinder. In order to prevent such derivation, an electronic force
to balance the Lorentz force is necessary. This is realized by two
parallel metal stripes wounded alternately with the two magnetic
stripes and connected to the cathode and the anode of a battery
respectively. The radial magnetic field component and the
electrical field along the axis of the cylinder are illustrated in
Fig.~\ref{fig1}(b). Thus the magnetic field in the cylinder takes
the form
\begin{eqnarray}
 \vec B ( z ) = B \vec n ( z  )\hspace{55mm} \nonumber \\
 = B\Bigl(\sin \vartheta \cos \bigl(\frac{2\pi }{L} z + \varphi \bigr),\,
        \sin \vartheta \sin \bigl(\frac{2\pi }{L} z + \varphi \bigr),\,
        \cos \vartheta
    \Bigr), \nonumber\\
\label{eq:B-field}
\end{eqnarray}
where $B=\sqrt {B_ \bot ^2  + B_z^2 }$,
$\vartheta=\tan^{-1}(B_\bot/B_z)$, $L$ is the length of the
cylinder, and $\varphi$ refers to the azimuthal angle of the
radial magnetic field component $\vec B_\bot$ at $z=0 $. Both $B$
and $\vartheta$ can be adjusted experimentally. Thus the magnetic
field is parameterized.

The magnetic field brings about a Zeeman energy to electrons
moving in the cylinder. Since the Lorentz force on the moving
electron is expected to be balanced by an extra-electric force as
aforementioned, the movement of the electron is governed by the
following Hamiltonian
\begin{equation}
H ( z ) = \frac{\mu}{2}\vec B\left( z \right) \cdot \vec \sigma,
\label{eq:hamiltonian}
\end{equation}
where $\vec \sigma =\left( \sigma _x,\, \sigma _y,\,\sigma _z
\right)$ denotes the Pauli operator and $\mu =g\mu _B$ the
coupling constant in which $g$ is the Land\'{e} factor and $\mu _B
= \frac{1}{2} \frac{e\hbar}{m}$ the Bohr magneton. Then the local
energy eigenstates of the Hamiltonian (\ref{eq:hamiltonian}) is
given by
\begin{eqnarray}
\left|  \uparrow _n\left(z\right) \right\rangle
= \cos \frac{\vartheta }{2}\left| \uparrow \right\rangle
 +e^{i\left( {\frac{{2\pi }}{L}z + \varphi } \right)}
 \sin \frac{\vartheta }{2}\left|  \downarrow  \right\rangle , \label{eq:local1}\\
 \left| \downarrow _n\left(z\right) \right\rangle
 = - \sin \frac{\vartheta }{2}\left| \uparrow \right\rangle
 + e^{i\left( {\frac{{2\pi }}{L}z + \varphi } \right)}
 \cos \frac{\vartheta }{2}\left|  \downarrow  \right\rangle .
\label{eq:local2}
\end{eqnarray}
where $\left | \uparrow \right\rangle$ and $ \left | \downarrow
\right\rangle $ refers to the two eigenstates of $\sigma_z$
operator. The eigenenergies of $|\uparrow_n\!(z)\rangle $ and
$|\downarrow_n\!(z)\rangle $  are $\mu B/2 $ and $-\mu B/2 $
respectively.

If an electron is travelling along the axis of the cylinder at
speed $v$, it will feel that the rotating frequency of the
magnetic field $\omega = 2\pi v/L$. We can control electrons'
speed $v$, let $\omega = \frac{{2\pi }}{L}v \ll \frac{\mu
B}{2\hbar}$, so that the adiabatic condition is satisfied. As the
magnetic field at $z=L$ returns to its original direction at
$z=0$, i.e. $\vec B \left ( L \right) =\vec B \left( 0 \right)$,
the evolution of the electron state is cyclic. According to the
adiabatic theorem in quantum mechanics \cite{berryphase}, the
local energy eigenstates (\ref{eq:local1}) and (\ref{eq:local2})
acquire extra phase factors in addition to the dynamical phase
factors:
\begin{eqnarray}
 \left|  \uparrow _n\left(L\right) \right\rangle &=& e^{i\delta } e^{i\gamma _B }
 \left|  \uparrow _n\left(0\right) \right\rangle , \\
 \left|  \downarrow _n\left(L\right) \right\rangle &=& e^{ - i\delta } e^{ - i\gamma _B }
 \left|  \downarrow _n\left(0\right) \right\rangle ,
 \end{eqnarray}
where
$\delta = \int_0^L \frac{\mu }{2\hbar v}B (z) dz $ is
the dynamical phase  and
$\gamma _B = \pi \cos \vartheta $
is the Berry phase.

Now we consider that electrons polarizing along the $z$ direction
are injected into the field. The incoming states may have $\left|
{\chi _ \uparrow  } \right\rangle _{{\rm{in}}}  = \left|  \uparrow
\right\rangle $ and $\left| {\chi _ \downarrow  } \right\rangle
_{{\rm{in}}}  = \left|  \downarrow  \right\rangle $. Below we
study their evolution in the adiabatically rotating magnetic
fields. At $z=0$, we expand the incoming states into the local
energy eigenstates:
\begin{eqnarray}
 \left| {\chi _ \uparrow  } \right\rangle _{{\rm{in}}} &=& \cos \frac{\vartheta }{2}\left|  \uparrow _n\left(0\right) \right\rangle - \sin \frac{\vartheta }{2}\left|  \downarrow _n\left(0\right) \right\rangle, \\
 \left| {\chi _ \downarrow  } \right\rangle _{{\rm{in}}} &=& e^{ - i \varphi } \left( \sin \frac{\vartheta }{2}\left|  \uparrow _n\left(0\right) \right\rangle + \cos \frac{\vartheta }{2}\left|  \downarrow _n\left(0\right) \right\rangle \right).
 \end{eqnarray}
Then the corresponding outgoing states for the electrons reaching
$z=L$ can be obtained by means of an unitary evolution:
\begin{widetext}
\begin{eqnarray}
 \left| {\chi _ \uparrow} \right\rangle _{{\rm{out1}}}
 &=&\Bigl( {\sin ^2 \frac{\vartheta }{2}e^{ - i\delta } e^{ - i\gamma _B }
 + \cos ^2 \frac{\vartheta }{2}e^{i\delta } e^{i\gamma _B } } \Bigr)
 \left|\uparrow \right\rangle  + \sin \frac{\vartheta }{2}\cos \frac{\vartheta }{2}e^{i\varphi }
  \Bigl( {e^{i\delta } e^{i\gamma _B }  - e^{ - i\delta } e^{ - i\gamma _B } } \Bigr)\left|
  \downarrow  \right\rangle,
      \label{eq:mup} \\
 \left| {\chi _ \downarrow} \right\rangle _{{\rm{out1}}}
 &=& \sin \frac{\vartheta }{2}\cos \frac{\vartheta }{2}e^{ - i\varphi}
 \Bigl( {e^{i\delta } e^{i\gamma _B }  - e^{ - i\delta } e^{ - i\gamma _B } }
 \Bigr)\left|  \uparrow  \right\rangle  + \left( {\sin ^2 \frac{\vartheta }{2}e^{i\delta }
 e^{i\gamma _B }  + \cos ^2 \frac{\vartheta }{2}e^{ - i\delta } e^{ - i\gamma _B } } \right)
 \left|  \downarrow  \right\rangle . \label{mdown}
 \end{eqnarray}
\end{widetext}
To guarantee the adiabatic condition being satisfied, the change
of the dynamical phase $\delta$ should be much larger than the
Berry phase shift $\gamma_B$ induced by the adiabatic excursion of
the system's parameters. Thus the separation of $\gamma _B$ from
the total phase requires very precise control of the system in
experiment\cite{rotation}.

As is known the spin-echo technique \cite{spinecho1, spinecho2} is
applicable for cancelling out the dynamical phase shift. We apply
a second setup similar to what we described in Fig. \ref{fig1} but
turning each magnet to the opposite direction and rotating the
cylinder by $\pi$. This provides a rotating magnetic field with
the opposite direction in comparison to the first one, i.e.,
$\vartheta\rightarrow\pi-\vartheta$ and $\varphi\rightarrow \pi +
\varphi$ in Eq. (\ref{eq:B-field}). Let the electron just passing
through the first setup inject into the second setup. Clearly, the
outgoing states (\ref{eq:mup}) and (\ref{mdown}) are the incoming
states with respect to the second setup. The outgoing states after
passing through the second cylinder are obtained as follows
\begin{widetext}
\begin{eqnarray}
\left| {\chi _ \uparrow  } \right\rangle _{{\rm{out2}}}
= \Bigl( \sin ^2\frac{\vartheta }{2}e^{ - 2i\gamma_B }
 +\cos ^2 \frac{\vartheta }{2}e^{2i\gamma _B } \Bigr)
   \left| \uparrow\right\rangle + \sin \frac{\vartheta }{2}\cos \frac{\vartheta}{2}e^{i\varphi }
\left( e^{2i\gamma _B}- e^{ -2i\gamma _B} \right)\left|\downarrow  \right\rangle ,
     \label{up}\\
\left| {\chi _ \downarrow  } \right\rangle _{{\rm{out2}}}= \sin
\frac{\vartheta}{2}\cos \frac{\vartheta }{2}e^{ - i\varphi }
\left( e^{2i\gamma _B}- e^{ -2i\gamma _B} \right)\left| \uparrow
\right\rangle  + \Bigl( \sin ^2 \frac{\vartheta }{2}e^{2i\gamma _B}
+ \cos ^2\frac{\vartheta }{2}e^{ -2i\gamma _B} \Bigr)
\left|\downarrow\right\rangle,
\label{down}
\end{eqnarray}
\end{widetext}
in which the dynamical phases are cancelled completely.
Eqs.~(\ref{up}) and (\ref{down}) exhibit that the rotating
magnetic fields alter the polarization direction of electron
spins. It therefore suggests a route for spin control
\cite{spintronics} merely by means of the justification of the
magnitudes of $\vec B_\bot$ and $\vec B_z$.
\begin{figure}[bph]
\vbox to 1.1in{\rule{0pt}{1.6in}} \includegraphics{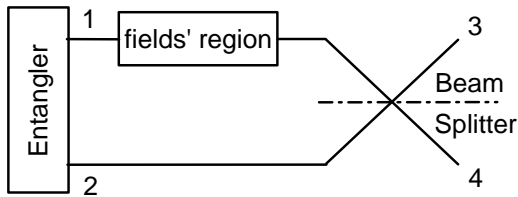} \caption{The
sketch picture of a 4-lead beam-splitter, in which two
adiabatically rotating magnetic fields are displaced in lead
1.}\label{fig2}
\end{figure}

Now we are in the position to study the transport properties of
electrons in an entangled system with one subsystem evolving in
adiabatically rotating magnetic fields. We suggest a scheme based
on the one presented in Ref.~\cite{setup, rashba}, but displacing
two afore-mentioned adiabatically rotating magnetic fields in one
incoming lead to construct `spin echo'~\cite{spinecho1,
spinecho2}; Fig.~\ref{fig2}. Entangled electron pairs are
generated by the entangler. Of each pair, one electron enters path
1 via lead 1 while the other enters path 2 via lead 2. We set the
entangler to keep injected electrons polarizing along the $z$
direction. Thus we have incoming states $\left|  \pm \right\rangle
_{\rm{in}} = \frac{1}{{\sqrt 2 }}\left( {\left| \uparrow
\right\rangle _1 \otimes \left| \downarrow \right\rangle _2 \pm
\left|  \downarrow \right\rangle _1  \otimes \left| \uparrow
\right\rangle _2 } \right)$. Because lead 1 is driven by the
adiabatically rotating magnetic fields, the incoming states will
be affected and change. However, there is no coupling between
paths 1 and 2. Hence the Hilbert space of the electron entangled
system is the direct product of two subspaces. It means that the
two subsystems evolve independently. The states of the subsystem 1
via path 1 undergoes an adiabatic evolution given by
Eqs.~(\ref{up}) and (\ref{down}), while those of the subsystem 2
via path 2 keeps their initial spin states. After passing through
the fields' region, we obtain the following outgoing states
%\begin{widetext}
\begin{eqnarray}\label{finalstate}
\ket{\pm }_{\rm out}
  =\frac{1}{\sqrt 2}\left[\cos^2\frac{\vartheta}{2} e^{2i\gamma_B}
   +\sin^2\frac{\vartheta}{2} e^{-2i\gamma_B}\right]\ket{\uparrow}_1 \otimes\ket{\downarrow}_2
        \nonumber \\
\pm\frac{1}{\sqrt 2}\left[\sin^2\frac{\vartheta}{2} e^{2i\gamma_B}
 +\cos^2\frac{\vartheta}{2}  e^{-2i\gamma_B}\right]\ket{\downarrow}_1 \otimes\ket{\uparrow}_2
          \nonumber \\
   + \frac{i\sin (2\gamma _B )}{\sqrt 2 }\sin\vartheta
   \Bigl(\,e^{i\varphi } \ket{\downarrow}_1 \otimes \ket{\downarrow}_2
   \pm e^{-i\varphi}\ket{\uparrow}_1 \otimes \ket{\uparrow}_2 \,\Bigr), \nonumber\\
\end{eqnarray}
%\end{widetext}
which is significantly affected by the Berry phase.

The outgoing electrons arrive at the beam-splitter and scatter. We
calculate shot noise for the scattering current in one outgoing
lead. Shot noise is regarded as a nonequilibrium current
fluctuation arising from the discrete nature of the charge flow
(at zero temperature) \cite{shotnoise,rashba}. The current
fluctuation around its mean value in lead $\mu$ at a time $t$ is
given by
$$\delta \hat{I}_\mu (t)=\hat{I}_\mu (t)- \langle \hat{I}_\mu (t) \rangle.$$
Usually, shot noise is defined as the Fourier transform of the
symmetrized current-current autocorrelation function between leads
$\mu$ and $\nu$:
\begin{equation} \label{noise}
S_{\mu \nu}(\omega)= \frac{1}{2}\int \langle \delta \hat{I}_\mu
(t) \delta \hat{I}_\nu (0) + \delta \hat{I}_\nu (0) \delta
\hat{I}_\mu (t) \rangle e^{i \omega t} dt.
\end{equation}
The current in lead $\mu$ in the scattering approach is
\begin{eqnarray}
\hat{I}_\mu (t) =\frac {e}{h} \sum _{\alpha \beta} \int d \epsilon d \epsilon'
 e^{i(\epsilon -\epsilon ') t/ \hbar} \mathbf{a}_\alpha ^\dagger (\epsilon)
  \mathbf{A}_{\alpha \beta} (\mu; \epsilon, \epsilon ') \mathbf{a}_\beta (\epsilon '),
   \nonumber\\
\end{eqnarray}
with the brevity notations:
\begin{eqnarray*}
\mathbf{A}_{\alpha \beta}(\mu; \epsilon, \epsilon ')&=&\delta _{\mu \alpha}
\delta_{\mu \beta} \mathbf{1}-\mathbf{s}_{\mu \alpha}^ \dagger
(\epsilon) \mathbf{s}_{\mu \beta} (\epsilon '), \\
\mathbf{a}_\alpha ^ \dagger & =& (a_ {\alpha \uparrow}
^\dagger, a_{\alpha \downarrow}^\dagger),
\end{eqnarray*}
where $a_{\alpha \sigma}^\dagger(\epsilon) [a_{\alpha
\sigma}(\epsilon)]$ denotes the creation (annihilation) fermionic
operator for an electron with spin $\sigma$ and energy $\epsilon$
in lead $\alpha$. The setup shown in Fig.~\ref{fig2} involves four
leads, and the single-particle scattering matrix elements to
describe it are $s_{31}=s_{42}=r$, and $s_{41}=s_{32}=t$, where
$r$ and $t$ denote the reflection and transmission amplitudes at
the beam splitter, respectively. We assume that there is no
backscattering, $s_{12}=s_{34}=s_{\alpha \alpha}=0$. The unitarity
of the scattering matrix implies $|r|^2+|t|^2=1$, and
$\mathrm{Re}|r^*t|=0$. At zero temperature, we calculate the shot
noise for the outgoing state (\ref{finalstate}) in the lead 3, and
find its zero-frequency component
%\begin{widetext}
\begin{equation}\label{shotnoise}
S_{33}^{\pm}  = \frac{{2e^2 }}{{h\nu }}T\left( {1 - T}
\right)f_{\pm}
\end{equation}
%\end{widetext}
where $T=|t|^2 $ stands for the beam-splitter transmission,
$\nu$ for the density of states in lead 3 which can
be determined by the mean value of the transport current $\left|
\left\langle {I_3 } \right\rangle \right|= \frac{e}{h\nu}$ and
\begin{eqnarray}
f_ +  &=& 1 - \delta _{\varepsilon _1 \varepsilon _2 } \left( {1 -
\frac{{2\gamma _B^2 }}{{\pi ^2 }}\sin ^2 \left( {2\gamma _B }
\right)} \right), \nonumber\\
f_ -   &=& 1 - \delta _{\varepsilon _1 \varepsilon _2 } \cos \left(
{4\gamma _B } \right) \nonumber
\end{eqnarray}
respectively for the states $\ket{+}_{\rm out}$ and $\ket{-}_{\rm out}$
with $\varepsilon_1$ and $\varepsilon_2$ the discrete energies of
the paired electrons.

\begin{figure}[bph]
\vbox to 2in{\rule{0pt}{2in}} \includegraphics{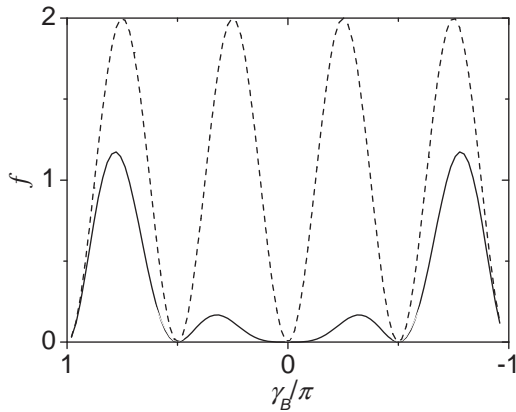}
\caption{Fano factor $f$ as a function of the Berry phase $\gamma
_B $. The solid line denotes the Fano factor for the state $\left|
+ \right\rangle _{{\rm{out}}} $ and the dashed line for the state
$\left|  -  \right\rangle _{{\rm{out}}} $.}\label{fig3}
\end{figure}

$f \equiv {S \mathord{\left/{\vphantom {S {\left[ {2eIT\left( {1 - T}\right)} \right]}}} \right.
 \kern-\nulldelimiterspace} {\left[ {2eIT\left( {1 - T} \right)} \right]}}
$ is called the Fano factor.
It has been shown ~\cite{setup} that the shot noise for the singlet state is noisy with an
enhanced Fano factor $2$, while that for triplets is noiseless
with a reduced Fano factor $0$. Hence the Fano factor for a
superstition of the singlet and entangled triplet evaluates
between $0$ and $2$.
The expression of Fano factors for states $\ket{\pm}_{\rm out}$ is distinct from those for
entangled states in Ref.~\cite{setup}, which implies that the Berry phase
modifies the spin symmetry of entangled states.

We plot the `normalized' Fano factor $f$ versus the Berry phase
$\gamma _B $ in Fig.~\ref{fig3}. Shot noises for
$\ket{\pm}_{\rm out}$ oscillate between bunching and antibunching,
but display different behaviors. For $\gamma _B =0$, the
difference in noise for $\ket{\pm}_{\rm out}$ and that for
entangled states vanishes. The relation between noise and the
Berry phase suggest a direct way to observe the Berry phase via
measuring shot noise in electron transport.

In conclusion, the Berry phase affects an electron's states
evolving in adiabatically rotating magnetic fields. Applying this
means, we change the entangled state into a linear superposition
of the singlet and triplets. The changed states are affected by
the Berry phase significantly, and the Fano factors for them in a
beam-splitter are single valued functions of the Berry phase, which
provides a new approach to detect the Berry phase in spin
transport on the basis of entanglement.

The work is supported by NSFC grant No.10225419, 10274069 \& 60471052, and
the Zhejiang Provincial Natural Foundation M603193.


\begin{thebibliography}{99}
\bibitem{scattertheory}M. B\"{u}ttiker, Phys. Rev. B 46, 12485 (1992).
\bibitem{beamsplitter}R. C. Liu et al., Nature (London) 391, 263
(1998).
\bibitem{beamsplitter2}M. Henny et al., Science 284, 296 (1999).
\bibitem{setup}G. Burkard et al., Phys. Rev. B 61, R16 303 (2000).
\bibitem{rashba}J. C. Egues, G. Burkard, and D. Loss, Phys. Rev. Lett. 89, 176401 (2002).
\bibitem{berryphase}M. V. Berry, Proc. R. Soc. London, Ser. A 392, 45 (1984).
\bibitem{book}Geometric Phase in Physics, edited by A. Shapere and
F. Wilczek (World Scientific, Singapore, 1989).
\bibitem{sjoqvist}E. Sj\"{o}qvist, Phys. Rev. A 62, 022109 (2000).
\bibitem{hessmo}B. Hessmo and E. Sj\"{o}qvist, Phys. Rev. A 62, 062301 (2000).
\bibitem{tong}D. M. Tong et al., J. Phys. A 36, 1149 (2003).
\bibitem{spinecho1}R. A. Bertlmann et al., Phys. Rev. A 69, 032112 (2004).
\bibitem{rotation}T. Bitter and D. Dubbers, Phys. Rev. Lett. 59, 251 (1987).
\bibitem{spinecho2}J. A. Jones et al., Nature (London) 403, 869 (2000).
\bibitem{spintronics}Semiconductor Spintronics and Quantum Computation, edited by D. D. Awschalom, D. Loss, and N. Samarth (Springer, Berlin, 2002).
\bibitem{shotnoise}Ya. M. Blanter and M. B\"{u}ttiker, Phys. Rep. 336, 1 (2000).

\end{thebibliography}
\end{document}